\documentclass[doublecol]{epl2} 
\usepackage{latexsym}
\usepackage{color}
\usepackage{bm}
\usepackage{amsmath}
\usepackage{amssymb}
\usepackage{ulem}
\usepackage{graphicx}

\title{The weak password problem: \\
chaos, criticality, and encrypted p-CAPTCHAs} 

\author{T.V.~Laptyeva\inst{1} \and S.~Flach\inst{1}\thanks{E-mail: \email{flach@pks.mpg.de}} \and K.~Kladko\inst{2}}
\shortauthor{T.V.~Laptyeva \etal}
 
\institute{                    
  \inst{1} Max-Planck-Institut f\"ur Physik komplexer Systeme - N\"othnitzer Stra\ss e 38, 
  D-01187 Dresden, Germany\\
  \inst{2} Axioma Research - 555 Bryant Street, Palo Alto, CA 94303, USA
}
 
\abstract{Vulnerabilities related to weak passwords are a pressing global economic and security issue.  
We report a novel, simple, and effective approach to address the weak password problem. Building upon chaotic dynamics, criticality at phase transitions, 
CAPTCHA recognition, and computational round-off errors we design an algorithm that strengthens security of passwords.  The core idea of our simple method is to 
split a long and secure password into two components. The first component is memorized by the user. The second component is transformed into a CAPTCHA image 
and then protected using evolution of a two-dimensional dynamical system close to a phase transition,  in such a way that standard brute-force attacks become 
ineffective. We expect our approach to have wide applications for authentication and encryption technologies.}

\begin{document}

\maketitle

\section{Introduction}
Computer and information security has been subject to intensive research for over 50 years. This included investigation of cryptographic methods, 
as well as generic security of computing devices, operating systems and networks. However, it is only relatively recently that the importance of the human 
factor has been given proper attention. Passwords are the common method for authentication and encryption used to secure digital life. Humans have 
limited capacity to remember passwords and tend to select passwords that are too simple and predictable. Security breaches related to weak passwords are 
widespread events. Consumers and enterprises around the world are looking for ways to address the weak password problem \cite{passwd1, passwd6}. In this 
paper we propose a simple method to address the problem by combining chaotic dynamics, phase transitions, and pattern recognition advantages of the human brain. 
We do not design a new encryption scheme. Instead we use standard encryption tschemes, and add a littel overhead on top in order to substantially enhance security.
A major building block of the proposed algorithm is the dynamic behavior of complex extended non-linear systems, in particular, Hamiltonian lattices close 
to a phase transition \cite{bose1, bose2}. These systems display non-ergodicity, deterministic chaos \cite{chaos}, and spontaneous formation of coherent 
space-time structures. Building upon dynamical chaos and computational round-off errors and utilizing superiority of the human brain over computers with 
respect to pattern recognition, our method protects a secret token, which can be used, in combination with a regular password, to derive a secret key for data encryption. 

It was estimated in 2009 that 86\% of US companies use password authentication and encryption \cite{zhang}.  A weak password used with a strong encryption or 
authentication algorithm potentially makes a computer system vulnerable to brute-force password search attacks. 
Studies have shown that users will generally address the password complexity problem by using simple predictable passwords \cite{as99,phnbpc09}. 
Schneier examined 34,000 MySpace online passwords and concluded that 65\% of them contained 8 characters, with most frequently used passwords 
being ``password1'', ``abc123'', ``myspace1'', and ``password'' \cite{phnbpc09}. Other user strategies include using the same password for every account, writing 
down passwords, storing passwords in files, and reusing or recycling old passwords. Horowitz reported that 15-20\% of the users on a regular basis wrote down their 
password on a Post-it note attached to the computer monitor \cite{phnbpc09}.  Another study found that 66\% of users keep password paper records at work and 58\% keep 
passwords in files \cite{phnbpc09}. 

Vulnerabilities related to weak passwords have significant economic effect globally. Results of a recent study \cite{mtm07,phnbpc09} revealed that identity fraud 
affects nearly 5\% of consumers, or nearly 10 million people in the USA per year. The total annual cost of identity fraud in the United States was more than \$55 billion in 
2006 \cite{mtm07}. Vulnerabilities related to weak passwords have significant economic effect globally. 

Cryptographic science utilizes discrete reversible functions that operate on bit strings and take a secret key as a parameter. As an example, Advanced Encryption Standard 
(AES) \cite{aes} specifies an encryption function approved for use by the US government. AES encrypts data in input/output blocks of 128 bits. The secret key lengths 
supported by AES are 128, 192, and 256 bits. These long key lengths were selected to make brute-force attacks infeasible. 

Over the years, a number of more sophisticated cryptanalysis attacks were described for various cryptographic algorithms. Such attacks are usually very technical and 
algorithm-specific and rely on finding statistical correlations in the cryptographic function to extract information on the cryptographic key. However, an ideal cryptographic 
function depends on its inputs in a completely random way with no correlations present. Therefore, a brute search attack remains the essential attack used in real world to 
compromise cryptographic algorithms.

For a completely random secret key used with the AES algorithm, a brute-force attack is presently infeasible and will probably remain so in the future. The situation
changes dramatically, when the key is limited to a 
smaller subspace of keys. A common situation is that the key is either a password, memorized by a human, or is derived from a password using a function known to the attacker. 
A brute search over a small subspace can then be done efficiently. 

A typical brute-force search attack requires that the attacker is in possession of the encrypted text (Ciphertext) C, and that the true key belongs to a subspace of keys S. 
The attacker can mount a Ciphertext-Only Attack by iterating through the space S and attempting to decrypt C in each case into a Candidate Plain Text. Now the attacker 
needs to determine whether it is the True Plain Text. This Recognition Problem is, therefore, a necessary part of Ciphertext-Only Attack, and amounts to designing an 
efficient algorithm denoted as the Recognition Oracle (RO). Implementations of ROs make use of the block-encryption structure, standard file formats, and correlations in True Plain Text.

\section{Proposed scheme}
We assume that a confidential data  file (D) is encrypted by a symmetric encryption algorithm, such as AES. 
 We also assume that the encryption and decryption are done by the same person, therefore we do not address vulnerabilities due to messenger capture attacks
when transmitting passwords. 
The encryption key EK is a combination of a reasonable-strength password 
component, 
which we denote as Short Password SP and an additional Strong Key SK:  EK=SP+SK. The difference between the proposed method and existing cryptographic 
technologies is that the user is not asked to memorize SK. Instead, the graphical representation of SK is embedded into a two-dimensional Image of 
Strong Key of 
a momentary initial state (IS)  of a nonlinear Hamiltonian two-dimensional lattice system. This embedding is similar to embeddings used in Completely Automated Public 
Turing test to tell 
Computers and Humans Apart (CAPTCHAs) \cite{captcha1,captcha2,captcha3}, therefore, we also coin it password CAPTCHA or p-CAPTCHA.
A time evolution of the two-dimensional lattice is then performed.  The chaotic evolution transforms the p-CAPTCHA into a chaotic final lattice state (FS).  Since our 
Hamiltonian system 
is close to a phase transition,
this chaotic state will contain regularities at various space scales, such as a domain structures. 
 Therefore the level of spatial correlations is the same both for IS and FS. 
If one encodes lattice site positions and velocities using floating point numbers, 
these regularities will be manifested in  the first half of the digits  (the more-significant bits) of such an encoding. 
 The second half of the digits  (the less-significant bits) will have a pseudo-random nature due to the dynamical chaos in the system.
\begin{figure}[htb]
\center{\includegraphics[width=0.8\columnwidth,keepaspectratio,clip]{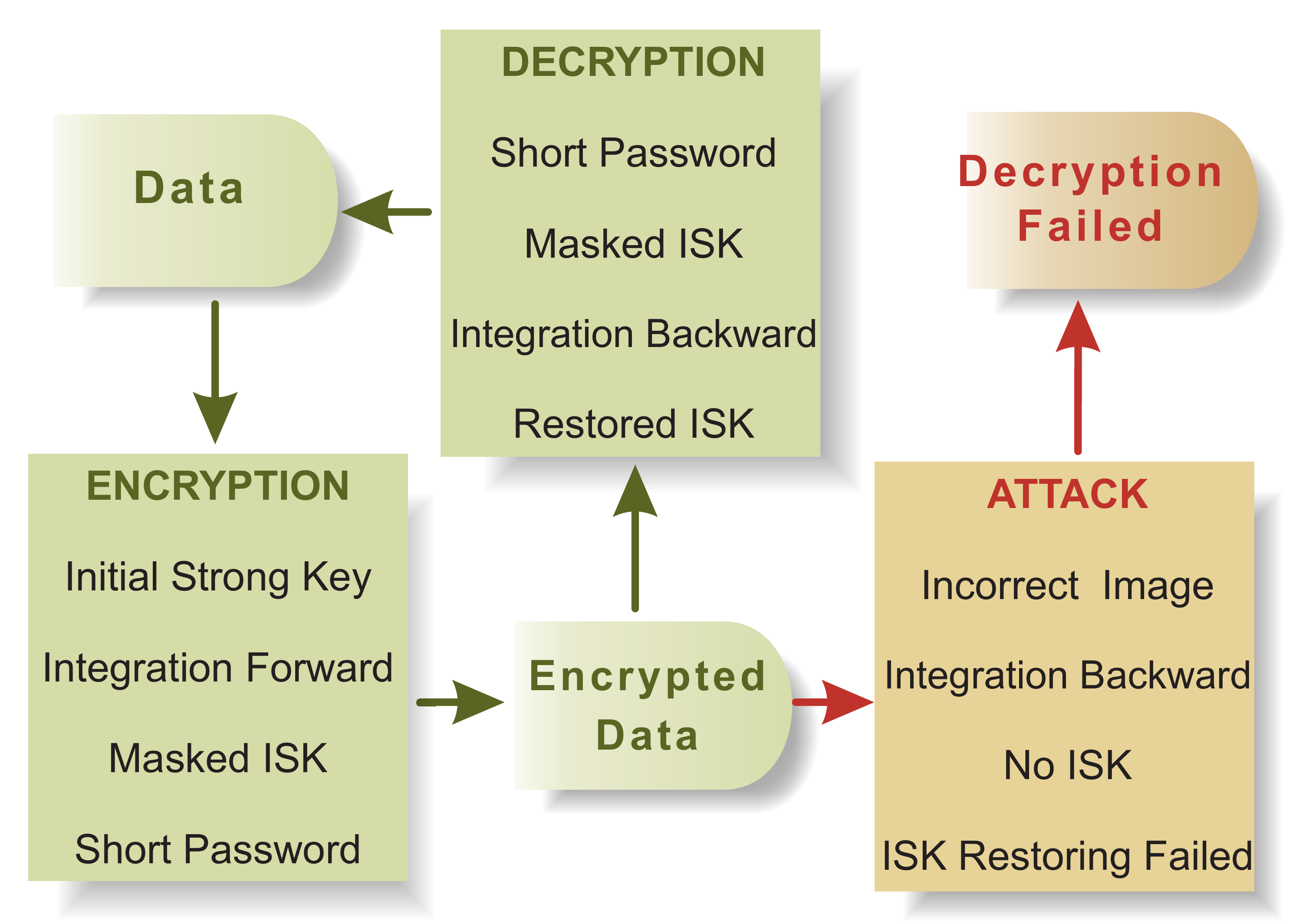}}
\caption{Schematic flow of the encryption, decryption and attack processes.}
\label{fig1}
\end{figure}
 
We split the state information of FS for each lattice site into two files. File F1 contains all more-significant bits, and file F2 contains all less-significant bits.
We encrypt F2 using the password SP, memorized by the user and obtain the encrypted file EF2. The scheme finally also encrypts the data D using the encryption key EK and
generates the encrypted data file ED. We glue all three files together into one resulting file EF=ED+F1+EF2.

To restore the data, EF is split into its three components ED, F1 and EF2. The user is then asked to enter SP in order to decrypt EF2 and to obtain F2. Files F1 and F2 
yield the correct final state FS. It is evolved back in time to the initial state IS. Its image is shown as a p-CAPTCHA. The user is asked to
read the characters and to type them in. The obtained strong key SK is combined with the already provided SP into the encryption key EK. Finally the encrypted data ED are
decrypted into the original data set D (see Fig.\ref{fig1}).

 Let us now assume that all three files ED, F1 and EF2 are available to the attacker.  
 
To mount a brute-force attack the attacker will
scan through the password space of SP. For each password the attacker will first try to decrypt EF2. 
Note that all (wrong) candidates for a decrypted F2 file will have the structure of a series of integers, therefore the attacker can not distinguish
wrong from right by checking their structure.
The attacker can now use such a candidate for F2, obtain a candidate for the final state FS, integrate backwards, and generate a candidate for the initial state IS.
The corresponding image will a random set of domains unless the correct SP was chosen initially. Since the dynamical system evolves at a fixed temperature
close to a phase transition, correlations of random domain wall images  and the p-CAPTCHA image are similar. The attacker is left with the option to
run an image recognition program over the candidate image. This takes 1-10 seconds (see discussion below) per recognition. 
%
 Therefore, computer-based Recognition Oracles will not be efficient  (see Fig.\ref{fig1}).

\section{Implementation}
\begin{figure}[htb]
\center{\includegraphics[width=0.8\columnwidth,keepaspectratio,clip]{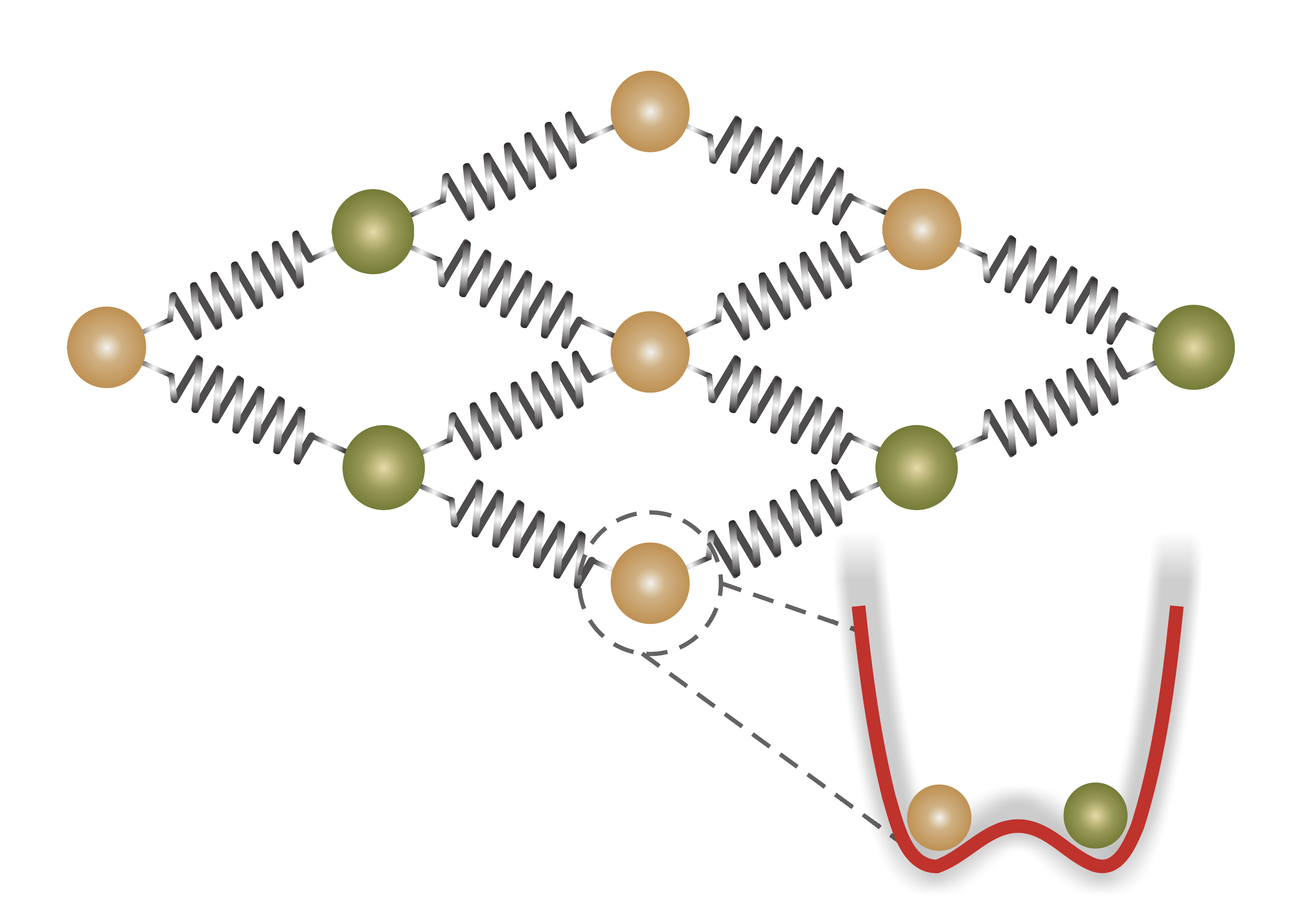}}
\caption{The two-dimensional square lattice of coupled double-well oscillators described by Eq.(\ref{eq:NKG}).
The springs indicate the nearest neighbour interactions. The double-well onsite potential for each oscillator includes 
two equilibrium positions $u_{ij}=\pm 1$.}
\label{fig2}
\end{figure}

In order to implement the strategy described above, we consider a two-dimensional square lattice of $N {{\times}} N$ coupled double-well oscillators depicted in 
Fig.\ref{fig2}, which is described by the Hamiltonian
\begin{eqnarray}
\mathcal{H} = \sum_{i,j=1}^N \left( \frac{1}{2}{p_{ij}^2}
-\frac{1}{2} u_{ij}^2+\frac{1}{4}u_{ij}^4 + \frac{1}{4}
+ \mathcal{F}_{ij} \right)\;, \nonumber \\
\mathcal{F}_{ij}=\sum_{k=\pm 1} \frac{1}{4} \left[ (u_{i+k,j}-u_{ij})^2 + (u_{i,j+k}-u_{ij})^2 \right]  \;.
\label{eq:NKG}
\end{eqnarray}
The lattice indices $i,j$ correspond to the two directions for the square lattice.
The equations of motion read $\dot{u}_{ij}=\partial \mathcal{H} / \partial p_{ij}$, $\dot{p}_{ij}=
-\partial \mathcal{H} / \partial u_{ij}$ and are invariant under time reversal.
We use $N=69$ and perform time evolution of the system using the symplectic Verlet algorithm \cite{lv67}.  The time step for the numeric integration is $h=0.01$, and double precision is used.
The system (\ref{eq:NKG}) served as a simple model for structural phase transitions e.g. in
ferroelectric materials as BaTiO$_3$ and also SrTiO$_3$ \cite{tses73}. 
The phase transition is of the second order \cite{hes71} at a certain critical value of the energy density 
which can be set roughly equal to the average temperature $T$. 
At high temperatures the oscillators traverse the potential barrier easily, therefore, 
the average polarization order parameter $M=\frac{1}{N^2}\left| \sum_{ij}{\rm sign}(u_{ij})\right|$ is zero for large $N$.
\begin{figure}[htb]
\center{\includegraphics[width=0.8\columnwidth,keepaspectratio,clip]{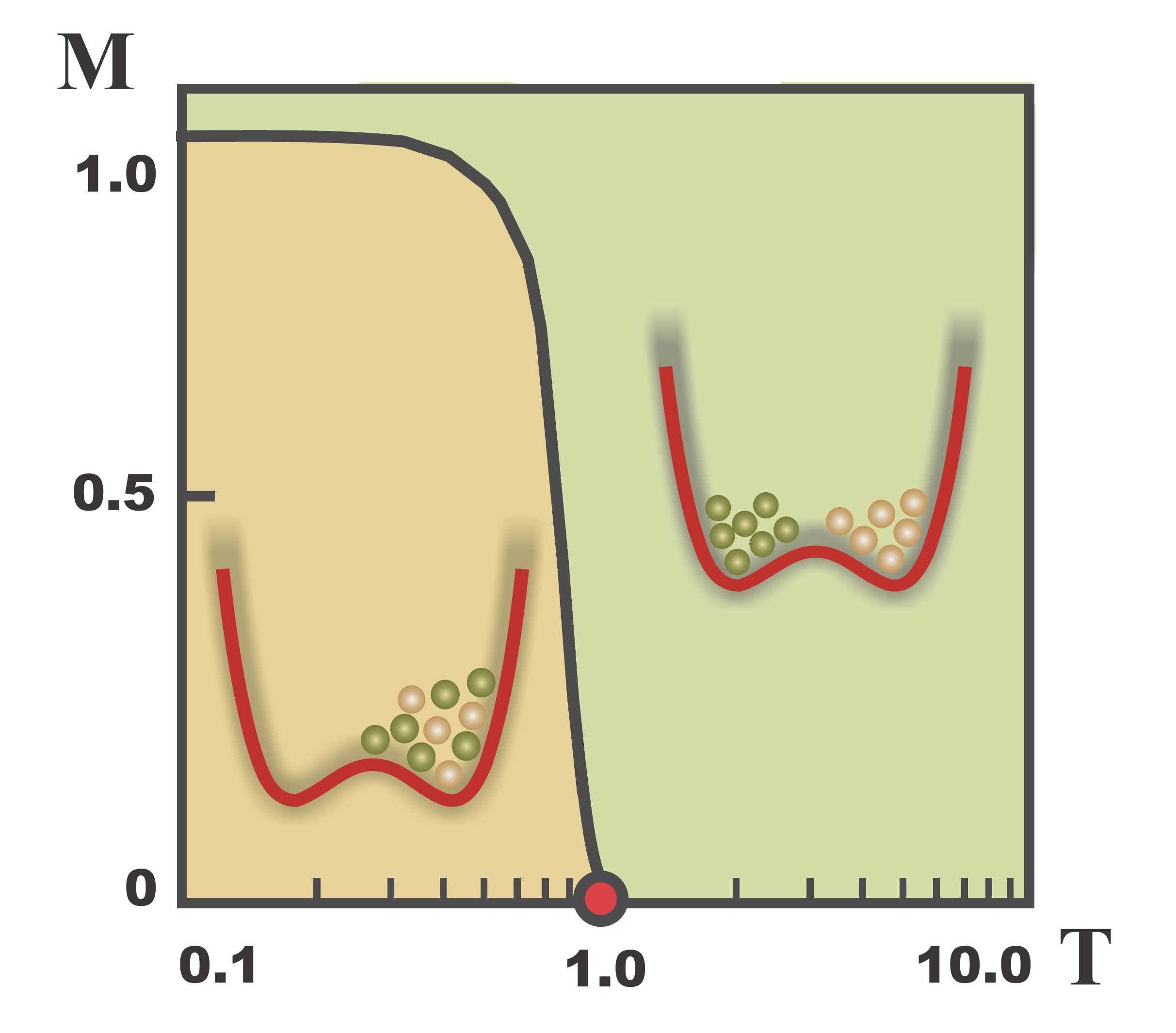}}
\caption{Dependence of the order parameter $M$ on the temperature $T$ for Eq.(\ref{eq:NKG}).
The red circle at the bottom indicates the operational point of the algorithm.}
\label{fig3}
\end{figure}
For low temperatures the energy of each oscillator is not sufficient to overcome the potential
barrier, and the interaction between oscillators enforces an ordered phase with $M \neq 0$.
The temperature dependence of $M$ is shown in Fig.~\ref{fig3}. The phase transition point is $T_c \approx 1$.

\begin{figure}[htb]
\center{\includegraphics[width=0.8\columnwidth,keepaspectratio,clip]{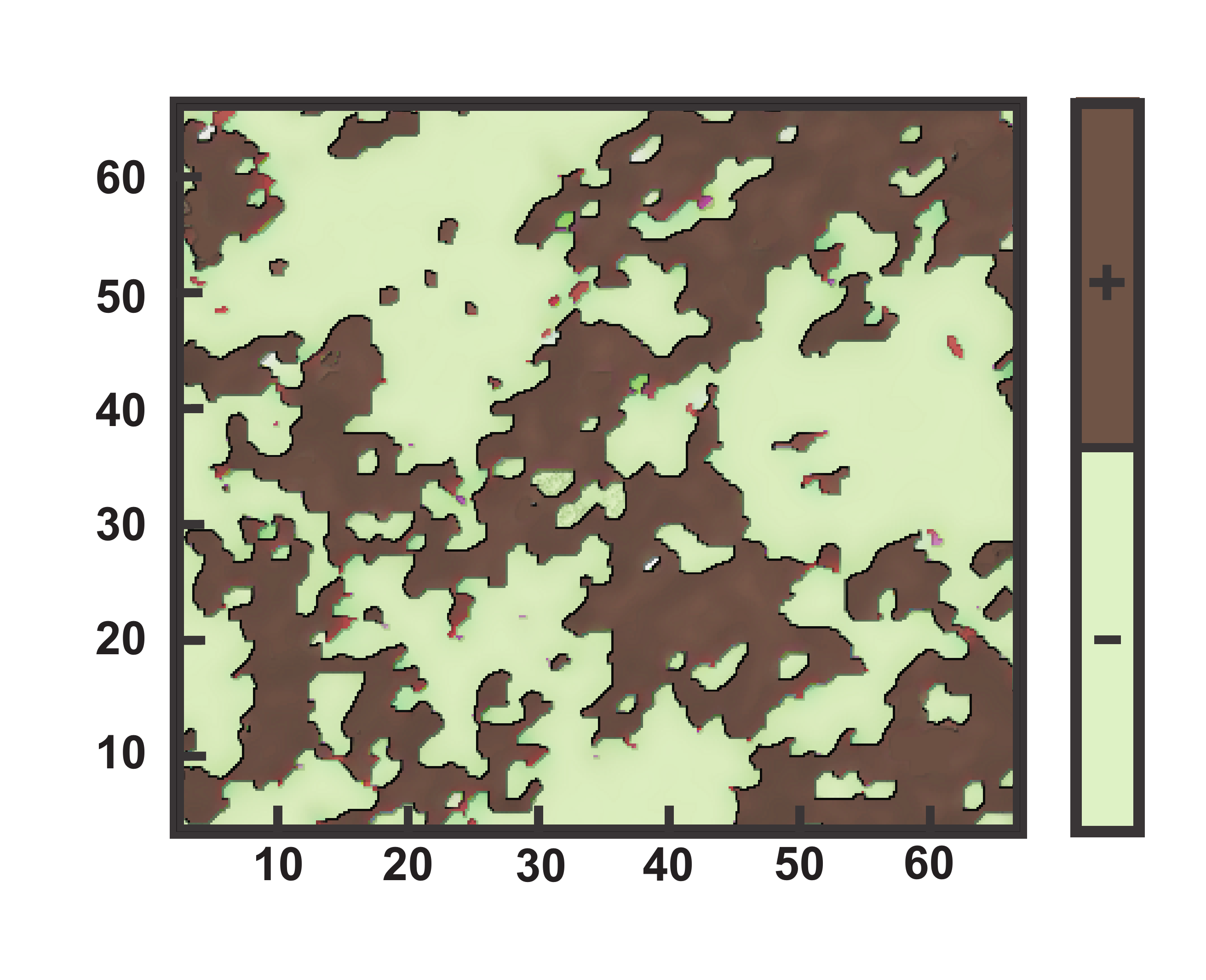}}
\caption{The thermalized state of Eq.(\ref{eq:NKG}) with parameters $T = 0.9$, $N=69$ in the color coding of coordinates after forward integration up to $\tau  = 200$ t.u.}
\label{fig4}
\end{figure}

The evolution of system (\ref{eq:NKG}) in the vicinity of the transition point is characterized by a spatial correlation length which diverges exactly at the phase transition 
point. Close to the phase transition large clusters of the low temperature phase emerge and disappear spontaneously. 
To initialize the system, we assign random values to the momenta $p_{ij}$ such that the kinetic energies $p_{ij}^2/2$ are distributed according to 
a Boltzmann distribution $\beta \rm{e}^{-\beta p^2/2}$ 
with a temperature $T\equiv \beta^{-1}=0.9$ (red circle in Fig.\ref{fig3}) and coordinates $u_{ij}=1$. We then integrate the equations of motion up to a time  of the order of $200$ 
time units (t.u.) at which all temporal correlations decay. The image of the thermalized local order parameter density distribution (the signs of the oscillator coordinates) is 
shown in Fig.~\ref{fig4}.

\begin{figure}[htb]
\center{\includegraphics[width=0.79\columnwidth,keepaspectratio,clip]{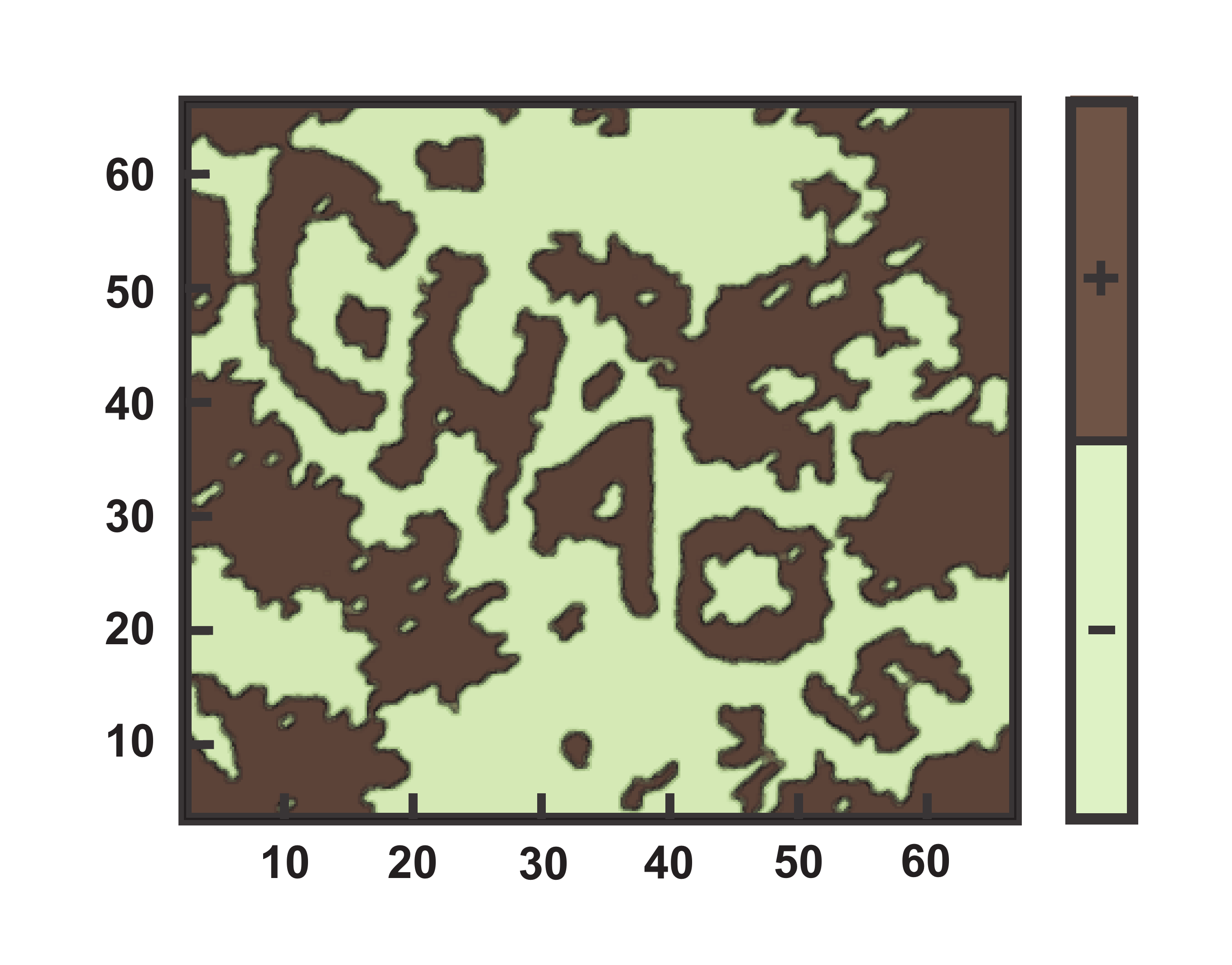}}
\caption{\textit{Initial} ISK (p-CAPTCHA) of Eq.(\ref{eq:NKG}) with parameters $T = 0.9$, $N=69$ in color coding of coordinates. Note that the \textit{restored} image 
(after forward integration to $\tau  = 350$ t.u. and backward integration to the origin) yields practically the same image.}
\label{fig5}
\end{figure}

\begin{figure}[htb]
\center{\includegraphics[width=0.8\columnwidth,keepaspectratio,clip]{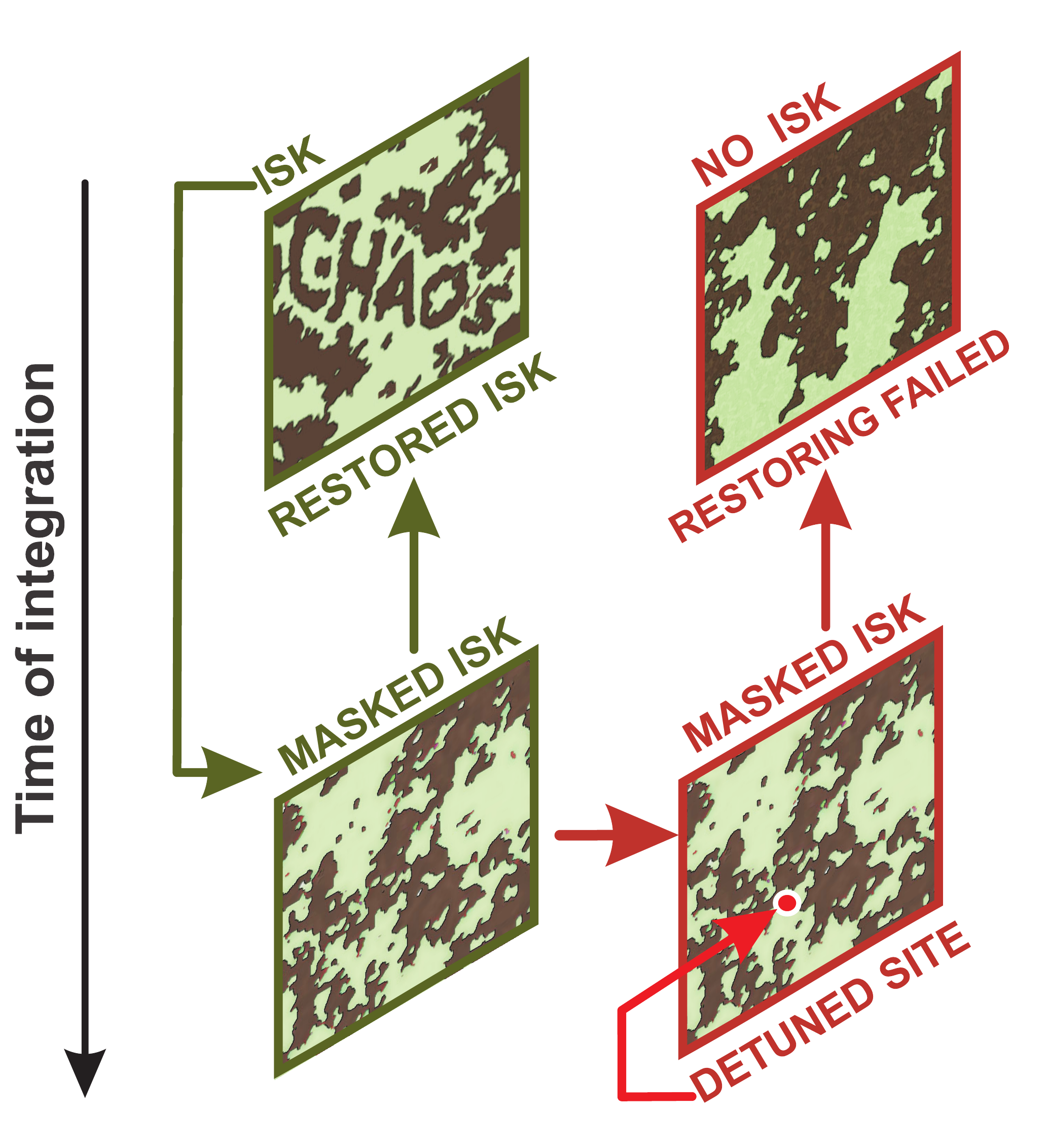}}
\caption{The evolution of the image p-CAPTCHA into a chaotic state and its reobtaining by integrating backwards (two left images). A slight detuning of one oscillator 
coordinate  $u_{20,20} \rightarrow u_{20,20}+0.00001$
(shown by the arrow in the bottom right image) of the chaotic state, followed by a backward integration, misses the image completely, leading to another random image 
of a chaotic state.}
\label{fig6}
\end{figure}

After that we imprint the SK (here the word ``CHAOS'') into the system and obtain 
the ISK or p-CAPTCHA (see Fig.~\ref{fig5} and left top image ``ISK/RESTORED ISK'' in Fig.\ref{fig6}). 
 The imprinting is done by using standard masks for deformed yet clearly visible keyboard characters. These masks are placed on the lattice, and the signs of all 
oscillator displacements inside such a mask area are set to '+', while keeping the absolute values of the displacements. The temperature is not affected by this operation. 

In order to protect SK, we integrate the equations of motion further to some time $\tau$ (bottom-left image ``MASKED ISK'' in Fig.\ref{fig6}).
Since the equations of motion are time reversible, we can invert the integration, and expect to
regain the original state ISK after back integration over the same time $\tau$. This is particularly true for the Verlet discretization, which is also completely time reversible. However,
the underlying dynamical system is non-integrable and, therefore, chaotic \cite{ajlmal82}.
Small perturbations will grow exponentially fast as $e^{\lambda t}$ where $\lambda$ is the
largest Lyapunov exponent \cite{ajlmal82}. We also note that the numerical integration algorithm, while being
perfectly invertible in time, generates round-off errors (for double precision, at the 15th digit
after the point). These small errors will accumulate exponentially fast in time. Therefore, there exists the maximum loopback time $\tau_\ast$ which still allows return to ISK. For larger loopback times the image ISK is lost in the high dimensional phase space of the system after the loopback evolution is performed. We find that $\tau_\ast \approx 400$ t.u.

Our strategy is then to choose $\tau$ to be close to $\tau_\ast$. With $\tau=350$ t.u. we can still integrate backwards and regain the image ISK. The restored image is 
practically identical to the original p-CAPTCHA we started with in Fig.\ref{fig5}.

Slightest errors in the velocities and positions of the oscillators will be amplified when integrating back,
and inhibit return to ISK. Indeed, we show this by slightly detuning the coordinate of an oscillator in the final state far from the original image location 
(right bottom image ``MASKED ISK WITH DETUNED SITE'' in Fig.\ref{fig6}):
$u_{20,20} \rightarrow u_{20,20}+0.00001$. Backward integration of the corrupted state leads to a loss of the ISK
(right top image ``NO ISK/RESTORING FAILED'' in Fig.\ref{fig6}).

\section{Benchmarking}

Let us discuss estimates for the security of the proposed algorithm. 
The knowledge of the password SP allows the legitimate user to decrypt the pseudo-random component and regain the correct state, which is then integrated in the reverse time direction,  
leading to IS. The strong key SK together with the short password SP is now used in a combination as a  secure secret token to decrypt protected data. 
 
Standard graphic cards - also called graphic processing unit (GPU) - are used by hackers to speed up brute force attacks. This is possible since GPUs contain
few hundreds of graphic processors (GP), as compared to one (sometimes two) central processing units (CPU) which are at the heart of each computer.
The difference is that GPs usually have only restricted memory as compared to CPUs, yet this will be not of importance here.
According to Honeyball \cite{honeyball} a login password with five characters is brute force cracked within one second (GPU) compared to 24 seconds (CPU).
Six character passwords need 4 seconds (GPU) or 90 minutes (CPU). Seven characters yield 18 minutes (GPU) versus 4 days (CPU). Finally, nine characters 
lead to 48 days (GPU) versus 43 years (CPU). The speedup factor in using GPUs is therefore of the order $3 \cdot 10^2$ in most cases, corresponding to
the number of GPs on the GPU.

We assume that we have about 80 different choices for one character in a given password. Short passwords with length $L$ will therefore span a space of
$80^L$ units. 
Our scheme requires CAPTCHA recognition. Automatic recognition programs need 1-10 seconds per image \cite{captcharecognition}.
Therefore, choosing the same length of a password as before, a user will gain additional protection with the presented method as
the {\sl additional} time needed for the brute force attack amounts to $80^L/1000$ seconds (note the the factor 1000 is already assuming that
all image recognition programs can be parallelized on the GPU). With this said, it is clear that our method gives additional security
with the same password length as used so far. However, we can even reduce the SP length, still keeping a high level of security. For instance,
a SP with $L=6$ amounts to a brute force attack with time 3000 years. Even a SP with length $L=5$ needs 38 days for a brute force attack to be successful.
To conclude these estimates, our scheme allows the user to memorize a five character password but have the protection of a standard nine character password,
with assuming that GPUs can be effectively used to perform image recognition.

\section{Outlook}
To conclude, we present an approach that relates the fields of dynamical chaos, criticality, and pattern recognition to cryptography. This approach allows us to use the 
evolution of a chaotic Hamiltonian system near a phase transition to embed and protect a secret token that can subsequently be used for cryptographic purposes, such as 
encryption of confidential data. Our method can be readily and straightforwardly implemented on a wide variety of existing computer systems and devices and, to our view, 
provides a significant step forward in protection of confidential data as compared to the currently available methods of password-based encryption. We hope that our findings 
can open a promising topic for future research. Potential future directions include searching for optimal Hamiltonian and non-Hamiltonian systems to be used as a foundation 
for our method, optimizing the performance of the method so that it can be executed on devices with low computational power, as well as designing better image embedding and 
evolution algorithms to provide stronger protections against computer-based image recognition.  

\acknowledgments
We thank P. Fulde and R. Khomeriki for useful discussions and a careful reading of the manuscript.

\end{document}